\documentclass{desyproc}

\begin{document}
\title{DarkSide-50: results from first argon run}

\author{{\slshape Davide D'Angelo for the DarkSide collaboration}\\[1ex]
Universit\`a degli Studi di Milano e I.N.F.N., via Celoria 16, 20133 Milano, Italy}

\contribID{315}

\confID{8648}  
\desyproc{DESY-PROC-2014-04}
\acronym{PANIC14} 
\doi  

\maketitle

\begin{abstract}
DarkSide (DS) at Gran Sasso underground laboratory is a direct dark matter search program based on TPCs with liquid argon from underground sources. The DS-50 TPC, with 50\,kg of liquid argon is installed inside active neutron and muon detectors. DS-50 has been taking data since Nov 2013, collecting more than $10^7$ events with atmospheric argon. This data represents an exposure to the largest background, beta decays of $^{39}$Ar, comparable to the full 3\,y run of DS-50 with underground argon. When analysed with a threshold that would give a sensitivity in the full run of about $10^{-45}$ cm$^2$ at a WIMP mass of 100\,GeV, there is no $^{39}$Ar background observed. We present the detector design and performance, the results from the atmospheric argon run and plans for an upscale to a multi-ton detector along with its sensitivity.
\end{abstract}

\begin{wrapfigure}{r}{0.30\textwidth}
\begin{centering}
\includegraphics[width=0.35\textwidth]{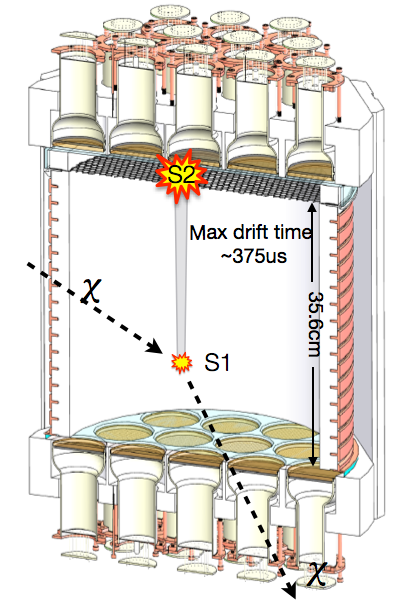}
\caption{DS-50 TPC principle of operation.}
\label{fig:tpc}
\end{centering}
\end{wrapfigure}

The DarkSide (DS) project \cite{Alexander:2013hia} aims to direct Dark Matter detection via WIMP-nucleus scattering in liquid Argon.
The detectors are dual phase Time Projection Chambers (TPCs) located at Laboratori Nazionali del Gran Sasso in central Italy under a rock coverage of $\sim$\,3800\,m\,w.e. 
DS aims to a background-free exposure via three key concepts: (1) very low intrinsic background levels, (2) discrimination of electron recoils and (3) active suppression of neutron background. 

DS has a multi-stage approach: after the operation of a 10\,kg detector \cite{Akimov:2012vv}, we are now running DarkSide-50 (DS-50) detector with a 45\,kg fiducial mass TPC and a projected sensitivity of $\sim$\,10$^{-45}$\,cm$^2$ for a 100\,GeV WIMP.
The project will continue with a multi-ton detector and a sensitivity improvement of two orders of magnitude.

The DS-50 TPC is depicted in Fig.~\ref{fig:tpc}. 
The scattering of WIMPs or background in the active volume induces a prompt scintillation light, called S1, and ionization. 
Electrons which do not recombine are drifted by an electric field applied along the z-axis. 
The maximum drift time across the 35.6\,cm height is $\sim$\,375\,$\mathrm{\mu}$s at the operative field of 200\,V/cm. 
Electrons are then extracted into gaseous argon above the extraction grid, where a secondary larger scintillation emission takes place, called S2. 
Two arrays of 19 3"-PMTs collect the light on each side of the TPC. 

\begin{wrapfigure}{r}{0.35\textwidth}
\begin{centering}
\includegraphics[width=0.35\textwidth]{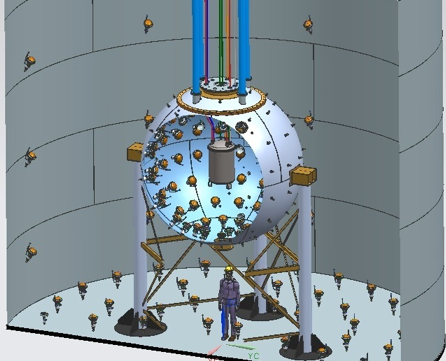}
\caption{DS-50 Schematics. TPC, ND and MD are visible.}
\label{fig:detector}
\end{centering}
\end{wrapfigure}

The TPC is housed inside an organic liquid scintillator Neutron Detector (ND) and a water Cherenkov Muon Detector (MD) \cite{PAGANI:2014ypa}, designed to host also a larger TPC with up to 5\,t of Liquid Argon, see Fig.~\ref{fig:detector}.
The ND is made by a 4\,m diameter steel sphere filled with a 1:1 mixture of Pseudocumene (PPO doped) and Trimethyl Borate (TMB) for enhanced neutron detection. 
The scintillation light is captured by 110 8"-PMTs mounted on the sphere inner surface. 
In addition of acting as a veto it also features independent trigger capabilities for an in-situ measurement of the neutron background. 
Boron has a high n-capture cross section which allows a compact veto size and reduces the capture time to 2.3\,$\mathrm{\mu}$s, two orders of magnitude below pure PC. 
The n-capture on $^{10}$B results in recoiling $^7$Li and $\alpha$ particle. 
In 94\% of the cases a 0.48\,MeV-$\gamma$ accompanies the process and is brightly visible. 
In the remaining cases the recoil energy of 1.47\,MeV must be detected and this is typically quenched to $\sim$\,50\,keV. 
Simulations indicate an efficiency $>$\,99\% for radiogenic neutrons and $>$\,95\% for cosmogenic neutrons \cite{Wright:2010bu}.   
The MD is a cylindrical tank, 11\,m in diameter and 10\,m high, filled with ultra-pure water and instrumented with 80 8" PMTs on the floor and inner walls. 
In addition of acting as water Cherenkov detectors for through-going muons with $>$99\% efficiency, it also serves as passive shielding again gammas and neutrons from the rocks.
DS-50 has been commissioned and is taking data since Nov. 2013.
After circulatimg Argon through charcoal filters for about 5 months, the electron lifetime was brought a stable value of $\sim$\,5\,ms, much larger then the maximum drift time in the TPC.

Operating Argon detectors implies dealing with the intrinsic cosmogenic background from $^{39}$Ar, a $\beta$-emitter with a Q=565\,keV and $\tau_{1/2}$\,=\,269\,y. 
In Atmospheric Argon (AAr) its activity can be as high as $\sim$\,1\,Bq/kg. 
However we have identified a source of Underground Argon (UAr) where the contamination is $<$\,6.5\,mBq/kg. 
A cryogenic distillation plant is producing the UAr at a rate of $\sim$\,0.5\,kg/d.
We are currently operating with AAr and we will switch to UAr at the beginning of 2015.
Argon has an intrinsic capability to distinguish Electron Recoils (ER) such as $^{39}$Ar decays from Nuclear Recoils (NR). 
Prompt scintillation light in Argon comes from the de-excitation of singlet and triplet states of Ar$_2^*$, having very different mean lives: $\tau_{\mathrm{singlet}}\sim$\,7\,ns while $\tau_{\mathrm{triplet}}\sim$\,1.6\,$\mathrm{\mu}$s.
Since NRs tend to populate more the singlet state, they result in significantly faster signals compared to ERs. 
We define the parameter F90 as the ratio of charge collected in the first 90ns over the total S1 charge. 
NRs are distributed around F90\,$\sim$\,0.7 while ERs around F90\,$\sim$\,0.3.

We have characterised our detector in terms of Light Yield (LY). 
At null field we have used the LY from the $^{39}$Ar shoulder at 565\,keV, obtaining LY$_{\mathrm{null}} \sim$\,8\,pe/keV, assumed energy independent within 3\%. 
With the application of the drift field, the LY becomes energy dependent and $^{39}$Ar is way beyond or region of interest. 
Therefore we spiked argon by adding gaseous $^{83m}$Kr in the recirculation system.  
$^{83m}$Kr decays fast ($\tau_{1/2} \sim$\,1.8\,h) and yields a good monochromatic line at 41.5\,keV. 
We have used the relative position of this line with and without drift field to scale the LY, obtaining LY$_{200V}\sim$\,7.2\,pe/keV at 200\,V/cm.

Compared to ERs, NRs are quenched by a factor that depends on energy and field. 
We have used the data from SCENE \cite{Cao:2014gns} to determine the quenching factor. 
SCENE features a small TPC with a concept similar to DS and has been measuring recoils from a neutron beam, whose energy can be selected.
SCENE has measured quenching factors at different neutron energies and drift field with respect to ERs from $^{83m}$Kr.
We have processed SCENE raw data using the DS reconstruction code and we have obtained the quenching factors as well as the distributions of the F90 parameter.

The ND has been also commissioned. 
The LY has been estimated exploiting $^{60}$Co contamination present in the cryostat. 
The LY has been found to be $\sim$\,0.5\,pe/keV, sufficient to detect recoils following a neutron capture on Boron of the order of 50\,keV$_{\mathrm{ee}}$. 
Unfortunately we have observed a high rate due to the intrinsic biogenic isotope $^{14}$C in the TMB, at the level of $\sim10^{-13}$\,g/g.
We have therefore successfully distilled the scintillator mixture and replaced TMB with pure Pseudocumene. 
Meanwhile we have identified a supplier of TMB coming from an underground oil batch which is low in $^{14}$C, $<10^{-15}$\,g/g. 
We will restore the design scintillator mixture before the end of 2014.

\begin{figure}[ht]
\begin{centering}
\includegraphics[width=0.570\textwidth]{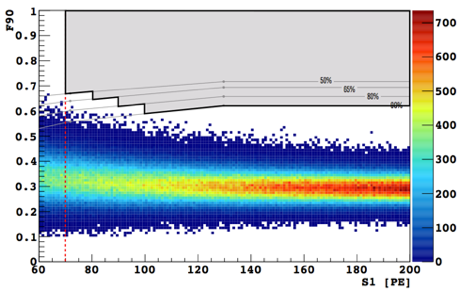}
\hfill
\includegraphics[width=0.38\textwidth]{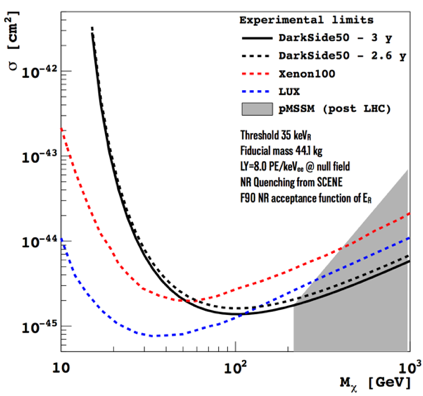}
\caption{DS-50 exposure of 280\,kg-days, F90 vs S1 energy in PE units with the NR acceptance curves and the WIMP search region superimposed (left). DS-50 projected sensitivity for 2.6\,y and 3\,y running with UAr compared to rejection curves from LUX and Xenon100 experiments (right).}
\label{fig:ds50}
\end{centering}
\end{figure}

In Fig.~\ref{fig:ds50} (left) are shown events corresponding to 280\,kg-days in the parameter plane of F90 vs S1 Energy in photoelectrons. 
Only single hit events are selected. 
A z-cut is applied to remove the regions close to the cathode and to the extraction grid. 
Events which show a coincident energy deposition in the ND are removed. 
The high $^{39}$Ar content of AAr  allows us to calibrate our S1-PSD with an exposure equivalent to 2.6\,y of operation with UAr at a contamination as high as the present upper limit. 
In this energy scale 70\,pe and 125\,pe correspond to $\sim$\,35\,keV and $\sim$\,57\,keV NRs according to the quenching factors determined from the SCENE data. 
70\,pe is also our choice of energy threshold. 
We have also superimposed the F90 NR acceptance curves derived from SCENE, a conservative choice as DS has a higher LY and hence narrower F90 distributions.
This plot proves that PSD at 200\,V/cm can efficiently suppress the dominant ER background that we expect in 2.6\,y of DS-50 UAr run, while maintaining high acceptance for WIMPs.
We have then proceeded to define a WIMP search region as in Fig.~\ref{fig:ds50}.
Assuming no candidate WIMP event, this allows us to project the sensitivity of DS-50 in the parameter plane of WIMP-nucleus cross-section vs WIMP mass, as it can be seen in Fig.~\ref{fig:ds50} (right), and compare it to the existing Xenon100 and LUX rejection curves.
Systematics on NR quenching factors and F90 curves contribute about 10\% variation at 100\,GeV WIMP mass.

\begin{wrapfigure}{r}{0.35\textwidth}
\begin{centering}
\includegraphics[width=0.35\textwidth]{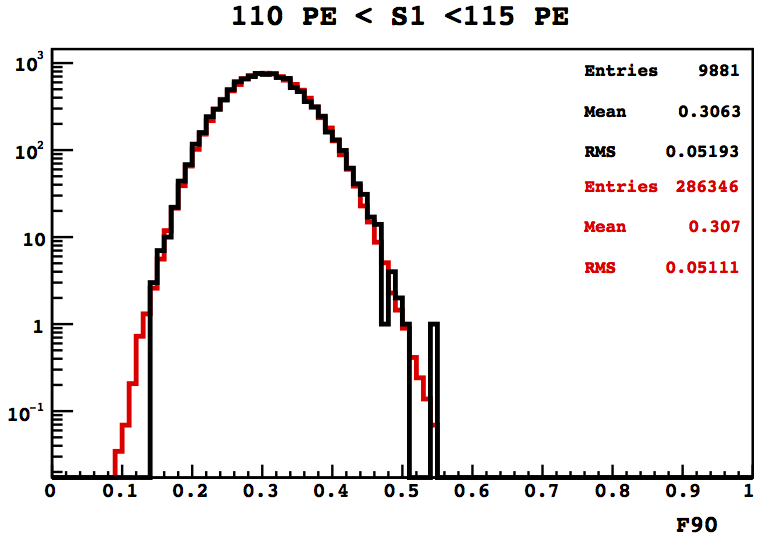}
\caption{F90 distributions in the [110,115]\,pe sample energy bin. Simulation (red) compared to data (black) after normalization.}
\label{fig:f90}
\end{centering}
\end{wrapfigure}

\begin{figure}[ht]
\begin{centering}
\includegraphics[width=0.53\textwidth]{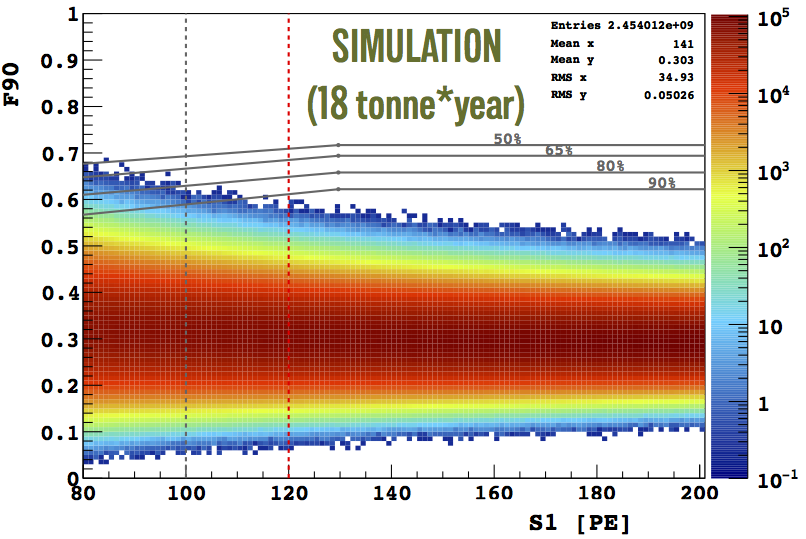}
\hfill
\includegraphics[width=0.42\textwidth]{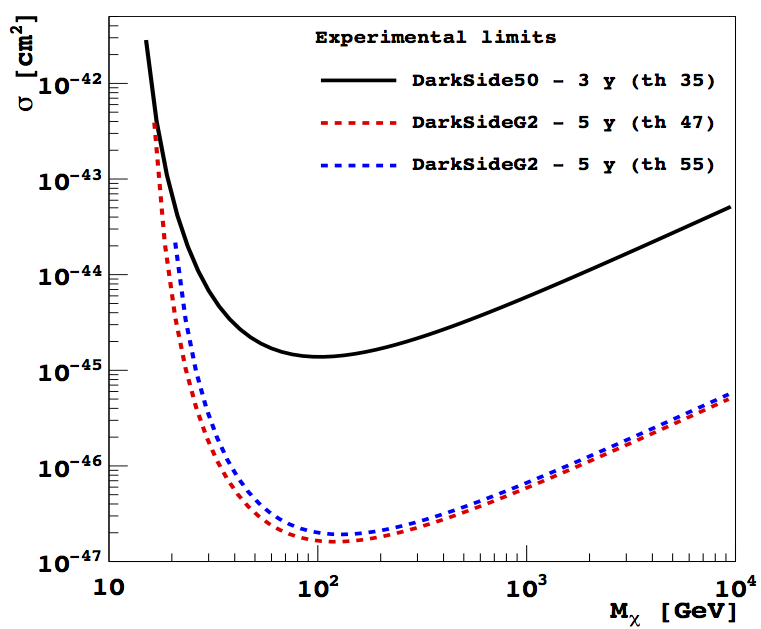}
\caption{DS multi-ton upscale. Simulated exposure of 5\,y, F90 vs S1 energy in PE units (left). Projected sensitivity with 120\,pe (blue) and 100\,pe (red) compared to DS-50 (right).}
\label{fig:dsG2}
\end{centering}
\end{figure}

We have also modelled F90 using the statistical distributions of the underlying processes with parameters taken from data. 
The model accounts for macroscopic effects related to argon micro-physics, detector properties, reconstruction and noise effects.
We have simulated F90 distributions for a DS upgrade of 3.8\,t fiducial mass and for 5\,y of run, assuming the ER background will be dominated by $^{39}$Ar at its present upper limit. 
Figure \ref{fig:f90} shows the agreement of the simulated distribution for a sample energy bin to real DS-50 data, after normalization. 
Similar plots are obtained for all energy slices.
Figure \ref{fig:dsG2} (left) shows the simulated exposure in analogy to the DS-50 data plot of Fig.~\ref{fig:ds50}. 
In this case the energy threshold would be 120\,pe although 100\,pe could be considered too. 
The projected sensitivity is shown in Fig.~\ref{fig:dsG2} (right). 
An increase in sensitivity of two orders of magnitude is expected in comparison with DS-50.

DS-50 has now acquired $\sim$\,5000\,kg-day of AAr data. The analysis is ongoing \cite{Agnes:2014bvk} in order to improve your understanding of backgrounds and study the S2 signal. The latter would bring x-y position reconstruction, hence a full 3D volume fiducialization, and additional ER/NR discrimination from the S2/S1 ratio.
We are also planning a detailed source calibration campaign in fall 2014.
In January 2015 we foresee to switch to UAr and start the physics run.


\begin{footnotesize}

\end{footnotesize}


\end{document}